\documentclass[prb,twocolumn,superscriptaddress,showpacs,aps]{revtex4}

\usepackage{color}
\usepackage{graphicx}
\usepackage{bm}
\usepackage{amsmath}

\begin{document}

\newcommand{\Tm}{TmNi$_2$B$_2$C}
\newcommand{\Lu}{LuNi$_2$B$_2$C}
\newcommand{\Y}{YNi$_2$B$_2$C}
\newcommand{\Tc}{T_{\text{c}}}
\newcommand{\Hci}{H_{\text{c1}}}
\newcommand{\Hcii}{H_{\text{c2}}}
\newcommand{\fq}{\phi_0}
\newcommand{\lamn}{\lambda_n}
\newcommand{\TN}{T_{\text{N}}}
\newcommand{\Jvec}{\textbf{J}}
\newcommand{\Hvec}{\textbf{H}}
\newcommand{\qvec}{\textbf{q}}
\newcommand{\rvec}{\textbf{r}}
\newcommand{\Bvec}{\textbf{B}}

\title{Small-angle neutron scattering study of the vortex lattice in superconducting {\Lu}}

\author{J. M. Densmore}
\author{P. Das}
\affiliation{Department of Physics, University of Notre Dame, Notre Dame, IN 46556}

\author{K. Rovira}
\altaffiliation{Department of Physics, Florida International University.}
%\altaffiliation[Permanent address: ]{Department of Physics, Florida International University, Miami, FL 33199}
\affiliation{Department of Physics, University of Notre Dame, Notre Dame, IN 46556}

\author{T. D. Blasius}
\altaffiliation{Physics Department, University of Michigan.}
%\altaffiliation[Permanent address: ]{Physics Department, University of Michigan, Ann Arbor, MI 48109}
\affiliation{Department of Physics, University of Notre Dame, Notre Dame, IN 46556}

\author{L. DeBeer-Schmitt}
\affiliation{Department of Physics, University of Notre Dame, Notre Dame, IN 46556}

\author{N. Jenkins}
\affiliation{DPMC, University of Geneva, 24 Quai E.-Ansermet, CH-1211 Gen\`{e}ve 4, Switzerland}

\author{D. McK. Paul}
\affiliation{Department of Physics, University of Warwick, Coventry CV4 7AL, United Kingdom}

\author{C. D. Dewhurst}
\affiliation{Institut Laue-Langevin, 6 Rue Jules Horowitz, F-38042 Grenoble, France}

\author{S. L. Bud'ko}
\author{P. C. Canfield}
\affiliation{Ames Laboratory and Department of Physics and Astronomy, Iowa State University, Ames, Iowa 50011, USA}

\author{M. R. Eskildsen}
\email{eskildsen@nd.edu}
\affiliation{Department of Physics, University of Notre Dame, Notre Dame, IN 46556}

\date{\today}

\begin{abstract}
We present studies of the magnetic field distribution around the vortices in {\Lu}. Small-angle neutron scattering
measurements of the vortex lattice (VL) in this material were extended to unprecedentedly large values of the scattering
vector $q$, obtained both by using high magnetic fields to decrease the VL spacing and by using higher order reflections.
A square VL, oriented with the nearest neighbor direction along the crystalline $[110]$ direction, was observed up to the
highest measured field.
The first-order VL form factor, $|F(q_{10})|$, was found to decrease exponentially with increasing magnetic field.
Measurements of the higher order form factors, $|F(q_{hk})|$, reveal a significant in-plane anisotropy and also allow for
a real-space reconstruction of the VL field distribution.
\end{abstract}


\maketitle

\section{Introduction}
%%%%%%%%%%%%%%%%%%%%%%%
The magnetic field distribution due to the vortex lattice (VL) in type-II superconductors depends on the detailed nature
of the superconducting state and on the properties of the host material. Examples of different field profiles are evident
if one considers the results of calculations based on different theoretical models for the superconducting
state.\cite{Abrikosov57,Ichioka99,Kealey00,Brandt03}
Experimentally one often seeks to parameterize the field modulation in terms of two characteristic length scales: the
penetration depth ($\lambda$) and the coherence length ($\xi$). Such an approach provides a simplified method of analyzing
the results of small-angle neutron scattering (SANS),\cite{Eskildsen97b} muon spin rotation ($\mu$SR)\cite{Price02} and
nuclear magnetic resonance (NMR) measurements. However, in addition to the simplification, such an approach also requires
the implicit acceptance of a particular theoretical model while, in many cases, violating its premises by for example using
the Ginzburg-Landau model to extract a field dependent penetration depth and coherence length.\cite{Landau07,Maisuradze09}

In this paper we will describe a more complete, model-independent analysis of SANS measurements of the VL in \Lu, extended
significantly beyond the first order Bragg reflection, which is customarily the only one measured. Measurements of a large
number of reflections allows for a real-space reconstruction of the VL magnetic field profile, which will be discussed
in relation to the significant in-plane anisotropy of this material caused by the Fermi
surface\cite{Rhee95,Kim95,Dugdale99,Starowicz08} and the anisotropic pairing in the superconducting
state.\cite{Yang00,Andreone01,Boaknin01,Maki02,Izawa02,MartinezSamper03,Raychaudhuri04,Bobrov05} 
To the best of our knowledge only very limited efforts have been undertaken in measuring higher order VL reflections, the
most notable exception being the work on Sr$_2$RuO$_4$by Kealey {\em et al.}\cite{Kealey00}

\section{Experimental details}
%%%%%%%%%%%%%%%%%%%%%%%
{\Lu} is a non-magnetic member of the rare-earth nickelborocarbide family of superconductors with a critical temperature of
$\Tc = 16.6$~K.\cite{Canfield98} The single crystal used in the SANS experiment was grown by a high temperature flux
method,\cite{Canfield01} using isotopically enriched $^{11}$B to reduce neutron absorption, and subsequently annealed to
improve quality and reduce vortex pinning.\cite{Miao02} The sample had a mass of $\sim 1$~g and a disc-like
crypto-morphology with the $c$ axis parallel to the thin direction.

The experiment was performed at the D11 SANS instrument at the Institut Laue-langevin. Incident neutrons with wavelength
$\lamn = 0.45$~nm and wavelength spread of $\Delta\lamn/\lamn = 10\%$ were used, and the VL diffraction pattern was
collected by a position sensitive detector.
Measurements were performed at 2~K in horizontal magnetic field between $0.5$ and 6~T, applied parallel to both the
crystalline $c$ axis and the incoming beam of neutrons. Two different magnetic field-temperature histories were employed:
Field cooled (FC) from a temperature above $\Tc$, and zero field cooled (ZFC) followed by an increase of the magnetic field
at 2~K.

\section{Results}
\begin{figure*}
  \includegraphics{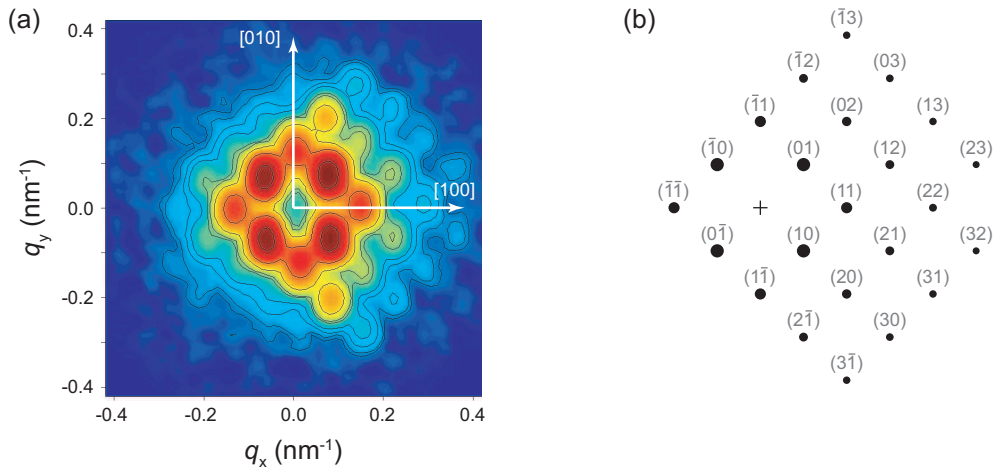}
  \caption{(Color online) SANS diffraction pattern of the VL in \Lu \ at $0.5$~T and 2~K following a field cooling
           procedure. The image (a) is a sum of measurements as the sample is rotated around the vertical axis in order to
           satisfy the Bragg condition for reflections in the center-right part of the detector. The data is smoothed and
           shown on a logarithmic scale.
           Measurements at 6~T, where no scattering from the VL could be observed, were used for background subtraction.
           The axes show the orientation of the crystalline axes.
           An indexing of the peaks is shown in (b), with the cross indicating the origin of reciprocal space.
           The apparent difference in intensity of e.g. the $(\bar{1}2)$ and $(2\bar{1})$ reflections compared to $(12)$
           and $(21)$ is due to different Lorentz factors, for which the detailed reflectivity analysis have been
           corrected.
           \label{DifPat}}
\end{figure*}

%%%%%%%%%%%%%%%%%%%%%%%
Here we present SANS imaging of the VL in \Lu \ to an unprecedentedly high field of $5.5$ T corresponding to 75\% of the
upper critical field, $\Hcii(2 \mbox{ K}) = 7.3$ T.\cite{Metlushko97,Shulga98,Budko01}
At all fields a square VL was observed as shown in Fig.~\ref{DifPat}(a).
An indexing of the VL Bragg reflections is shown in Fig.~\ref{DifPat}(b). The different reflections are located at a
distance from the center of the detector which is proportional to their scattering vector,
$q_{hk} = (h^2 + k^2)^{1/2} \, q_0$; where $q_0 = 2 \pi (B/\fq)^{1/2}$ and $\fq = h/2e = 2070$~Tnm$^2$ is the flux quantum.
With increasing field, the VL Bragg peaks move out in reciprocal space and their intensities decrease, and as a consequence
fewer peaks are visible. At 5~T and above only the $\{10\}$-reflections are observed. Measurements performed at 6~T were
used for background subtraction. While this is below $\Hcii$ no scattering from the VL could be observed at this field.
Furthermore, the detailed measurement in Fig.~\ref{FFhkvsQ} show that at $q_{10}(6 \mbox{~T}) = 0.34$~nm$^{-1}$ the
extrapolated (10) intensity is more than an order of magnitude smaller than that of any observed reflection at lower fields
at the same $q$.

\subsection{Vortex lattice symmetry and orientation}
\label{VLsym}
At low applied magnetic fields the VL in \Lu \ undergoes a field-driven symmetry- and reorientation
transition.\cite{Eskildsen97a,Levett02} This two-step transition arises due to the growing importance of the Fermi surface
anisotropy coupled with non-local electrodynamics as the vortex density increases.\cite{Kogan97}
At higher fields, it has been proposed theoretically that thermal vortex fluctuations may lead to a re-entrance of the
square VL phase.\cite{Gurevich01}
A similar re-entrance was also predicted by Nakai {\em et al.} who considered a case where competing superconducting gap
and Fermi surface anisotropies both favor a square VL, but oriented at $45^{\circ}$ with respect to one
another.\cite{Nakai02} In addition to the re-entrance of the square VL phase stabilized by the Fermi surface anisotropy,
this model predicts a $45^{\circ}$ rotated square VL phase at even higher fields due to the gap anisotropy.

Experimentally we found that above 5~T the VL reflections broaden significantly in the azimuthal direction. While this
could be due to the onset of reentrance of the square VL phase (transition back into a rhombic
symmetry)\cite{Eskildsen01,Dewhurst05} no splitting into two peaks was seen, and therefore the broadening may also simply
be due to a disordering of the vortex lattice. No indication of a $45^{\circ}$ VL rotation was observed. Whereas this
does not exclude such a transition at even higher fields it significantly reduces the fraction of the $HT$-phase diagram
where it can occur. This result therefore imposes significant constraints on the model parameters used in the calculations
described above.

\subsection{Vortex lattice reflectivity and form factor}
\begin{figure}
  \includegraphics{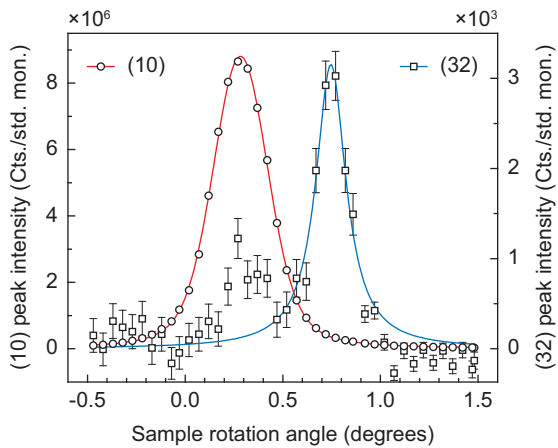}
  \caption{(Color online) Rocking curves at $0.5$~T and 2~K for the \Lu \ $(10)$ and $(32)$ VL reflections from
           Fig.~\ref{DifPat}. Note the different intensity scales for the two reflections. Error bars for the (10)
           reflection are not shown since they are smaller than the size of the data points. The intensity at each angular
           setting are obtained by summing the detector counts at the position of the specific Bragg reflection. The curves
           are Voigt fits to the data. The shoulder seen for the $(32)$ reflection is unrelated to the VL, as discussed in
           the text.
           \label{RC}}
\end{figure}

Measuring the intensity of the VL reflections as the sample is rotated around the vertical axis to satisfy the Bragg
condition, provides rocking curves as the ones shown in Fig~\ref{RC}.
In addition to the strongest $(10)$ reflection, the figure shows the $(32)$ rocking curve which was the highest order
reflection visible at a field of $0.5$~T. The intensity of these two reflections differ by a factor of 3000.
The longer scattering vector for the $(32)$ reflection, $q_{32} = \sqrt{13} \, q_0$, is evident by the larger rotation
angle necessary to satisfy the Bragg condition.
To obtain the VL reflectivity, the integrated intensity was determined by fitting a Voigt function to each rocking curve
and normalizing the area to the incident beam intensity. Compared to other functional forms (e.g. Gaussian or Lorentzian)
the Voigt was found to provide a significantly better fit to the data. The difference in the Lorentz factor (angle between
the scattering vector and the vertical rotation axis) for the two reflections gives rise to a difference in the width of
the two rocking curves, as they are cutting through the Ewald sphere at different incident angles. The integrated intensity
for all reflections is corrected for this effect.
What ultimately limits how many VL Bragg peaks can be imaged is the vanishing intensity and imperfect background
subtraction as seen for the (32)-reflection in Fig.~\ref{RC}. Here significant background variation is clearly evident,
even leading to an apparent shoulder on the VL rocking curve. While this could be interpreted as being due to a second
VL domain, this is clearly not the case since a similar shoulder is not seen on the (10) rocking curve.

With the strong VL peaks in \Lu, especially at low fields as shown in Fig.~\ref{DifPat}, it is necessary to consider
whether multiple scattering is affecting the measured intensities. Four multiple scattering processes can affect the
measured intensity of a given Bragg reflection:
(i) Extinction by scattering back into the incident beam, 
(ii) a diminished incident beam intensity due to scattering into other reflections,
(iii) scattering into other reflections (``aufhellung''), and
(iv) scattering from other reflections (``umweganregung'').\cite{Moon64}
The first 3 processes decrease the intensity of the specific reflection whereas the last one increases it. Furthermore for
(ii) to (iv) to occur multiple VL reflections must satisfy the Bragg scattering condition simultaneously. Since the effects
of multiple scattering depend on the magnitude of the reflectivity, the (10), (01) and (11) reflections, which peak at the
same sample rotation angle, are the most likely to be affected. Following Moon and Shull,\cite{Moon64} and taking into
account the simplification presented by the SANS geometry, one finds that to leading order the corrections due to multiple
scattering are $P_{10}/P_0 = R_{10} + R_{10} \, R_{11}$ and $P_{11}/P_0 = R_{11} + R_{10}^2$.
Here $P_{10}$ and $P_{11}$ are the measured scattered power, $P_0$ is the measured power of the transmitted incident beam,
and $R_{10}$ and $R_{11}$ are the intrinsic VL reflectivities. Taking the peak intensity of the (10) reflection at $0.5$~T
from Fig.~\ref{RC} ($\approx 9 \times 10^6$ cts./std. mon.) yields $P_{10}/P_0 = 0.3\%$. Using the measured
$P_{11}/P_{10} = (0.6)^2 \approx 0.4$ (see Fig.~\ref{FFratiovsH} below) and the above expressions, we find that the
difference between the normalized scattering and the reflectivity are less than 1\% for both the (10) and (11) reflections
and thus insignificant compared to the typical error of $20\%$ with which the scattering powers can be measured.

The reflectivity is proportional to the square modulus of the VL form factor $F(q_{hk})$, which is the Fourier transform
at wave vector $q_{hk}$ of the two-dimensional magnetic flux modulation of the VL. The reflectivity and the form factor
for a given reflection is related by
\begin{equation}
  R_{hk} = \frac{2 \pi \gamma^2 \lamn^2 t}{16 \fq^2 q_{hk}} \left| F(q_{hk}) \right| ^2,
  \label{Refl}
\end{equation}
where $\gamma = 1.91$ is the neutron gyromagnetic ratio, $t$ is the sample thickness, and $q_{hk}$ is the magnitude of the
scattering vector.\cite{Christen77}
Since the vortex spacing, and consequently $q$, depends on the magnetic field ($q \propto \surd H$), this allows $|F(q)|$
to be determined continuously over wide range of scattering vectors using SANS.
In the following Section we will focus on the field dependence of the first-order VL form factor, $|F(q_{10})|^2$, and in
Section \ref{Fhk} on the higher order form factors, $|F(q_{hk})|^2$.

\subsection{Field dependence of $|F(q_{10})|$}
\label{F10}
\begin{figure}
  \includegraphics{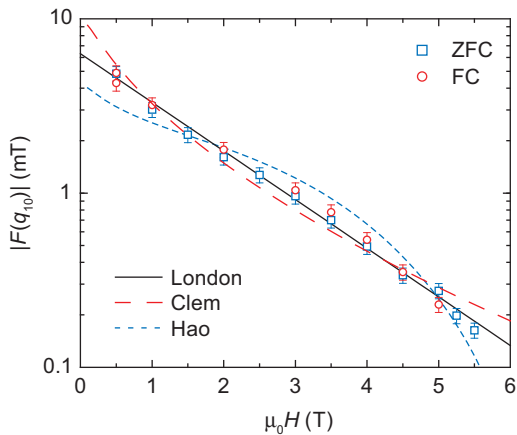}
  \caption{(Color online) Field dependence of the VL $(10)$ form factor for both the field cooled (FC) and zero field
           cooled (ZFC) case. The fitted values of the penetration depth and coherence lengths are
           $\lambda = 90.7$~nm and $= 8.22$~nm for the London model ($\chi^2 = 0.14$),\cite{Eskildsen97b,Yaouanc97}
           $\lambda = 61.9$~nm and $\xi = 12.7$~nm for the Clem model ($\chi^2 = 0.62$),\cite{Clem75} and
           $\lambda = 104.1$~nm and $\xi = 7.19$~nm for the Hao model ($\chi^2 = 1.20$).\cite{Clem75b,Hao91}
           \label{FF10vsH}}
\end{figure}

Using the integrated intensity obtained from rocking curves, such as the ones shown in Fig.~\ref{RC}, and utilizing
Eq.~(\ref{Refl}), one obtains the field dependence of the VL form factor of the $(10)$-reflections shown in
Fig.~\ref{FF10vsH}.
Within the experimental error, field cooling and zero field cooling produce identical results; this is indicative of very
low pining in the sample which is due in part to the post-growth annealing.\cite{Miao02}

Several models exist for the form factor field dependence. By far the simplest model is based on the London model, extended
by a Gaussian cut-off to take into account the finite extent of the vortex cores:\cite{Eskildsen97b,Yaouanc97}
\begin{equation}
  F(q) = \frac{B}{1 + (\lambda q)^2} \; e^{-c (\xi q)^2}.
  \label{London}
\end{equation}
Here $\lambda$ and $\xi$ are respectively the penetration depth and coherence length, and the constant $c$ is typically
taken to be between $1/4$ and 2.\cite{Yaouanc97} As shown by the solid line in Fig.~\ref{FF10vsH}, the measured form factor
is well fitted by this model which corresponds to a simple exponential decrease with increasing field. Since for all the
fields applied $(\lambda q)^2 \gg 1$ the prefactor in Eq.~(\ref{London}) reduces to $\fq/(2\pi \lambda)^2$. This is in
agreement with our earlier results,\cite{Eskildsen97b} but here extended to significantly higher fields. Using $c = 1/2$
the exponential fit to the form factor yields $\lambda = 90.7$~nm and $\xi = 8.22$~nm. This value for the coherence length
is in excellent agreement with previous results. The penetration depth is about $15\%$ shorter,\cite{Eskildsen97b} which is
consistent with an improvement of the sample quality by annealing. The value of $\xi$ is also in reasonable agreement with
the estimate based on the upper critical field, $\xi_{\text{c2}} = 6.7$~nm.

A more rigorous model for the form factor field dependence was obtained by Clem by including an effective core radius
$\xi$, and solving the Ginzburg-Landau model.\cite{Clem75} This was later extended by Hao and Clem to include the
suppression of the bulk order parameter due to vortex overlap.\cite{Clem75b,Hao91} Fits to both of these models are shown
in Fig.~\ref{FF10vsH}. The most noticeable difference between the two models is the s-shaped form factor of the Hao model,
and the significant downturn at higher fields due to the proximity to the upper critical field $\Hcii$. What is also clear
from Fig.~\ref{FF10vsH} is that the form factor is somewhat better described by the London model ($\chi^2 = 0.14$) compared
to either the Clem ($\chi^2 = 0.62$) or Hao ($\chi^2 = 1.20$) models, and furthermore that the the Clem model returns
unrealistic values for $\lambda$ and $\xi$. In this regard, it should also be pointed out that the Hao model has in
general been shown to be a poor approximation to exact, numerical solutions of the Ginzburg-Landau
model.\cite{Brandt03,Brandt97}
The field dependence of the VL form factor in \Lu, which is commonly considered to be a
relatively simple superconductor, emphasize the point that any analysis of bulk measurements based on a particular
theoretical model for the VL must be done with the utmost care. Finally, it is clear that measurements of $F(q_{10})$ alone
provides limited insight into the VL field distribution.

\subsection{Higher order form factors, $|F(q_{hk})|$}
\label{Fhk}
\begin{figure}
  \includegraphics{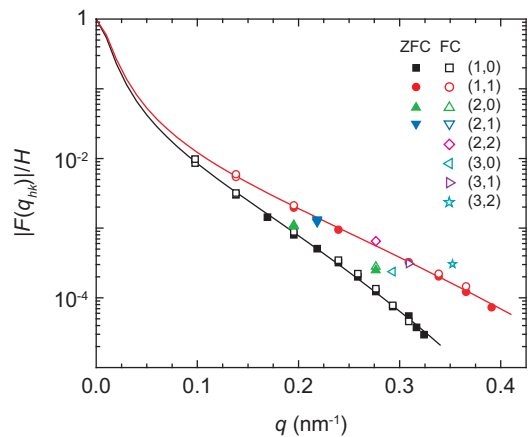}
  \caption{(Color online) VL form factor divided by the applied field versus scattering vector $q$ for all measured
           reflections. Curves through $|F(q_{10})|$ and $|F(q_{11})|$ are fits to the London model (Eq. (\ref{London}) and
           Fig.~\ref{FF10vsH}).
           \label{FFhkvsQ}}
\end{figure}

The VL form factors for all measured reflections and fields are summarized in Fig.~\ref{FFhkvsQ}.
In both the London and Clem model the only field dependence of $|F(q_{hk})|/H$ comes through the magnitude of the
scattering vector, $q_{hk}$, and therefore the form factors would be expected to collapse onto a single curve. This is not
observed for \Lu; instead, the form factor follows a different exponential field dependence, as shown for $|F(q_{10})|/H$
and $|F(q_{11})|/H$.
Likewise the data does not agree with the Hao model, which predicts that for a given $q$ the form factor should increase
with decreasing field (larger indices $h$ and $k$) and converge towards the value given by the Clem model and observed in
niobium.\cite{Clem75b,Hao91} Rather we observe that while $|F(q_{11})|/H$ does indeed lie above $|F(q_{10})|/H$, other
higher order form factors fall in-between these two limiting curves.

\begin{figure}
  \includegraphics{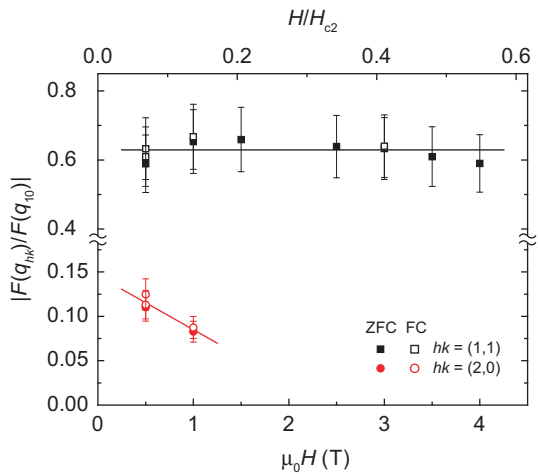}
  \caption{(Color online) Field dependence of VL form factor ratios $F(q_{11})/F(q_{10})$ (top) and $F(q_{20})/F(q_{10})$
           (bottom). Note different scales. The black line is the average and the red line is a fit to the data.
           \label{FFratiovsH}}
\end{figure}

The deviation from the theoretical predictions is also evident if one considers the field dependence of the form factor
ratio $|F(q_{11})/F(q_{10})|$ shown in Fig.~\ref{FFratiovsH}.
At a field of $0.5$~T, all the models predict a value of $\sim 0.3$, decreasing monotonically by a factor between two and
four as the field in increased to 4~T. It should however be noted that numerical calculations for a square VL in an
$s$-wave superconductor predict a largely field independent value of $|F(q_{11})/F(q_{10})|$,\cite{Ichioka99} which is in
agreement with the experimental results presented here.

The failure of the theoretical models to describe the measured form factors is not surprising when one keeps in mind that
they were all derived assuming the screening current plane to be isotropic. It is well known that \Lu, as well as
the other members of the rare earth nickelborocarbide superconductors, posseses a significant in-plane
anisotropy.\cite{Rhee95,Kim95,Dugdale99,Starowicz08,Yang00,Andreone01,Boaknin01,Maki02,Izawa02,MartinezSamper03,
Raychaudhuri04,Bobrov05}
The four-fold, in-plane anisotropy manifests itself most strongly in the experimental data in Fig.~\ref{FFhkvsQ}, when
comparing the $[110]$ and $[100]$ crystalline directions, corresponding to the $(10)$ and $(11)$ VL-reflections.

\section{Discussion}
%%%%%%%%%%%%%%%%%%%%%%%
While it would be straightforward to incorporate an in-plane penetration depth anisotropy into the models discussed above,
our emphasis here will be on a model-free determination of the VL field modulation.

\subsection{Real-space field reconstruction}
\begin{figure*}
  \includegraphics{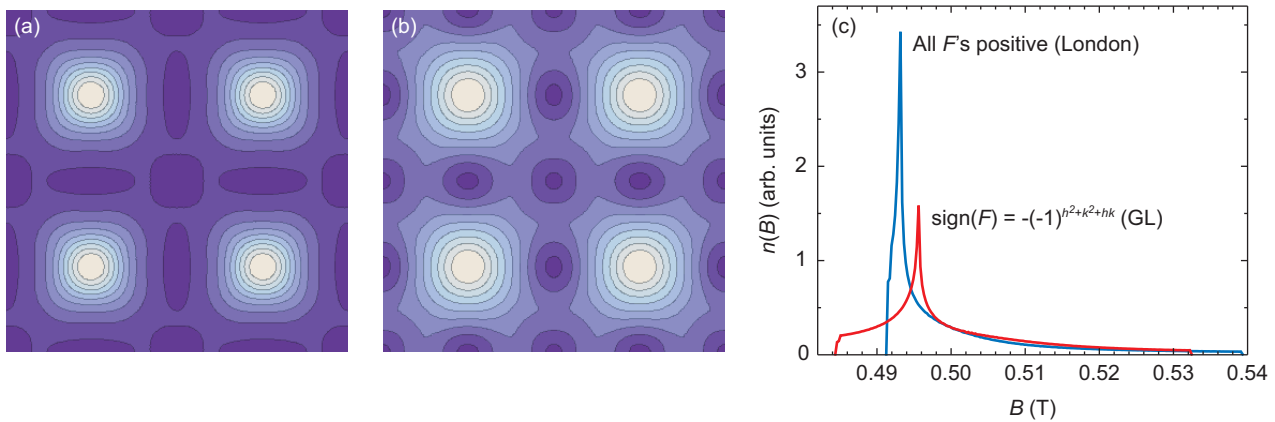}
  \caption{(Color online) Real space magnetic field reconstruction from the measured VL form factors at $0.5$~T and 2~K,
           using the London (a) and Ginzburg-Landau (b) sign schemes.
           In both cases an equidistant contour spacing of 5 mT was used, with the lowest contour at respectively
           493~mT~(a) and 486~mT~(b). In both cases the amplitude of the field modulation is $48.9$~mT or roughly 10\% of
           the average (applied) magnetic field.
           Four VL unit cells are shown, with a vortex spacing of $64.3$~nm corresponding to an applied field of
           $\mu_0 H = 0.5$~T. As discussed in the text the true field reconstruction is with all form factors positive (a).
           Note that the field reconstruction images are rotated $45^{\circ}$ with respect to Fig.~\ref{DifPat}, so that
           the $\{110\}$-directions are now horizontal/vertical.
           Panel (c) shows the magnetic field distribution function for the two different field reconstructions.
           \label{FieldReconsmuSR}}
\end{figure*}

With the VL form factor being simply the Fourier transform of the magnetic field modulation, the real space field
distribution can be obtained from the measured form factors by
\begin{equation}
  B(\bm{r}) = \sum_{hk} F(\bm{q}_{hk}) \: e^{i \bm{q}_{hk} \cdot \bm{r}}.
  \label{Bsum}
\end{equation}
In the case of the VL the so-called phase problem, arising from the fact that only the magnitude $|F(q_{hk})|$ is measured,
is greatly simplified. As the magnetic field variation around any vortex exhibits inversion symmetry
($B(-\bm{r}) = B(\bm{r})$), the form factor must be real and the phase problem thus reduces to a sign problem. Within the
London model the sign on all form factors is expected to the be the same, which will be chosen as positive and corresponds
to having the vortex at the center of the unit cell. In contrast, the Ginzburg-Landau model predicts both positive and
negative signs for the form factor, determined by the indices $h$ and $k$ according to
$-(-1)^{h^2 + k^2 + hk}$.\cite{Brandt7475} Note that $F(q_{00})$ is simply the applied magnetic field $\mu_0 H$. As shown
in Fig.~\ref{FieldReconsmuSR}(a,b) the field reconstruction obtained from Eq.~(\ref{Bsum}) differs significantly depending
on which sign scheme is used.
Given that the measurements were performed at a temperature and field much below both $\Tc$ and $\Hcii$, one would expect
the London sign scheme to apply. This is supported by comparing the field dependence of $|F(q_{20})/F(q_{10})|$, shown in
Fig.~\ref{FFratiovsH}, to the numerical work (Fig.~10) by M. Ichioka {\em et al.} [\onlinecite{Ichioka99}]. This shows how
the ratio is expected to decrease with increasing field, with the sign change separating the London from the
Ginzburg-Landau regime occuring at $H \approx 1/2 \, \Hcii$ for their choice of model parameters. In comparison, the ratio
$|F(q_{20})/F(q_{10})|$ in Fig.~\ref{FFratiovsH} extrapolates to zero at $H = 1/3 \, \Hcii$. Further support for the choice
of all positive form factor signs is obtained by calculating the magnetic field distribution function shown in
Fig.~\ref{FieldReconsmuSR}(c), and comparing the results to muon spin rotation ($\mu$SR) experiments.\cite{Price02}
Finally, using the field distribution ($B_{\text{peak}} - B_{\text{min}} = 1.79$~mT) and Eqs.~(12) and (13) from
ref.~\onlinecite{Maisuradze09} yields an estimate for the penetration depth $\lambda = 88.6$~nm, in excellent agreement
with our fit to the London model.

\subsection{Basal plane anisotropy}
As stated earlier \Lu \ possesses a substantial basal plane anisotropy arising both from the Fermi
surface\cite{Rhee95,Kim95,Dugdale99,Starowicz08} as well as the superconducting
pairing.\cite{Yang00,Andreone01,Boaknin01,Maki02,Izawa02,MartinezSamper03,Raychaudhuri04,Bobrov05}
Thermal conductivity measurements indicates a gap minima, or possibly even nodes, along
$\langle 100 \rangle$,\cite{Izawa02} leading to theoretical speculations of a ($s+g$) pairing symmetry.\cite{Maki02} Still
the dominating anisotropy appears to be due to the Fermi surface, as indicated by the square VL configuration and discussed
in Section~\ref{VLsym}.

\begin{figure}[b]
  \includegraphics{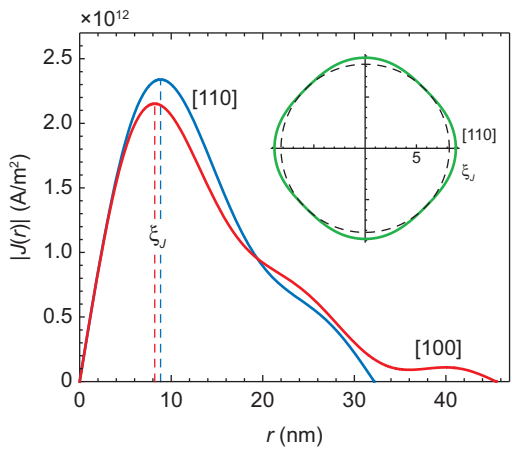}
  \caption{(Color online) Supercurrent density as a function of distance from the vortex center, along the VL nearest
           neighbor vortex direction (crystalline [110] axis) and the VL diagonal (crystalline [100] axis).
           The inset shows the value of $\xi_J$ (distance of maximum current) in the basal plane. To visually emphasize
           the four-fold anisotropy, a circle with radius $\xi_J^{[100]}$ is also shown (dashed line).
           \label{Jprofile}}
\end{figure}

A measure of the in-plane anisotropy can be obtained from the SANS results by simply calculating the current flow around
the vortices from the field reconstruction in Fig.~\ref{FieldReconsmuSR}(a) using $\mu_0 \bm{J} = \nabla \times \bm{B}$.
Fig.~\ref{Jprofile} shows $|\bm{J}(\bm{r})|$ along the VL nearest neighbor direction as well as the VL unit cell diagonal.
In cases like this, one frequently uses an operational definition of the coherence length ($\xi_J$) as the distance from
the vortex center to the maximum current.\cite{Ichioka99,Sonier04} From Fig.~\ref{Jprofile} it is clear that $\xi_J$
differs for the two directions shown. The inset to Fig.~\ref{Jprofile} shows $\xi_J$ in the basal (screening current)
plane. It is striking that the minimum $\xi_J$ is observed along the nodal $\langle 100 \rangle$-directions, where one
would naively expect the weakest pairing and hence the largest $\xi_J$.
It is important to emphasize that the field reconstruction is robust in the sense that the anisotropy of $\xi_J$ does not
change (although the absolute numerical values do) if one varies the measured form factors even well beyond the typical
10\% experimental error of their values. Rather, our result follows directly from the unusually large ratio
$|F(q_{11})/F(q_{10})|$ in \Lu. Finally we also note that the $\xi_J$ variation cannot be explained by a simple
``squeezing'' effect,\cite{Sonier04} since the value along the nearest neighbor direction ($[110]$) is larger than that
along the VL diagonal ($[100]$).

We speculate that our result may share a common origin with recent scanning tunneling microscopy measurements on
iso-structural \Y. These measurements showed a fourfold-symmetric star shaped vortex core extending in the
$\langle 100 \rangle$-directions at zero energy but splitting into 4 peaks and effectively rotating the vortex shape by
45$^{\circ}$ at higher energies within the superconducting gap.\cite{Nishimori04}

It is interesting to compare the field reconstruction and current profiles in Figs.~\ref{FieldReconsmuSR}(a) and
\ref{Jprofile} to the numerical work of Machida and Ichioka {\em et al.}, who have performed calculations for a number of
VL configurations, pairing symmetries and Fermi surface anisotropies.\cite{Ichioka99,MachidaPC}
From this it is clear that the best agreement is achieved with an anisotropic superconducting gap ($d$-wave or anisotropic
$s$-wave) combined with a Fermi surface anisotropy. Further numerical work to optimize the agreement between the
experimental and calculated results should provide valuable input to calculations of VL configuration, such as the one in
Ref.~\onlinecite{Nakai02}, and result in a more realistic VL phase diagram for \Lu.

\section{Summary}
%%%%%%%%%%%%%%%%%%%%%%%
To summarize we have performed comprehensive SANS measurements of the VL in \Lu, thus ending the common but unsatisfactory
practice of discarding all but the (1,0) reflection.
The measurements confirmed the existence of a square VL up to 75\% of $\Hcii$. 
The first-order VL form factor, $|F(q_{10})|$, was found to decrease exponentially with increasing magnetic field, in
agreement with the generalized London model but not with the supposedly more realistic models for the VL field
distribution.
Measurements of higher order form factors, $F(q_{hk})$, and the real-space reconstruction of the VL field modulation,
provide a qualitative measure of the in-plane anisotropy. This will therefore serve as important input to future
theoretical work.

Similar measurement and analysis should be performed on other members of the nickelborocarbides; indeed on any other
superconductor where enough higher order reflections are measurable. In this regard, the results presented here will serve
as a reference for future work.

\begin{acknowledgments}
%%%%%%%%%%%%%%%%%%%%%%%
We are grateful to Kazushige Machida, Masanori Ichioka and Vladimir Kogan for stimulating discussions, and to Hazuki
Kawano-Furukawa and Seiko Ohira-Kawamura for discussing their data on YNi$_2$B$_2$C with us prior to publication.

This work is supported by the National Science Foundation through grant DMR-0804887 (J.M.D. and M.R.E) and
PHY-0552843 (K.R. and T.D.B.).
M.R.E. acknowledges support by the Alfred P. Sloan Foundation.
Work at the Ames Laboratory was supported by the Department of Energy, Basic Energy Sciences under Contract No.
DE-AC02-07CH11358.
\end{acknowledgments}

% Bibliography
%%%%%%%%%%%%%%%%%

\newpage

\printfigures

\end{document}